\documentclass[showpacs,twocolumn,pre]{revtex4}
\usepackage{amsmath}
\usepackage{latexsym}
\usepackage{epsfig}
\usepackage{makeidx}
\usepackage{amsfonts}

\begin{document}

\title{Subdiffusive master equation with space dependent anomalous exponent:
`Black Swan' effects}
\author{Sergei Fedotov and Steven Falconer}
\affiliation{School of Mathematics, The University of Manchester, Manchester M60 1QD, UK }

\begin{abstract}
We derive the fractional master equation with space dependent anomalous
exponent. We analyze the asymptotic behavior of corresponding lattice model
both analytically and by Monte Carlo simulation. We show that the
subdiffusive fractional equations with constant anomalous exponent $\mu $ in
a bounded domain $\left[ 0,L\right] $ are not structurally stable with
respect to the non-homogeneous variations of parameter $\mu $. In
particular, the Gibbs-Boltzmann distribution is no longer the stationary
solution of the fractional Fokker-Planck equation whatever the space
variation of the exponent might be. We analyze the random distribution of $%
\mu$ in space and find that in the long time limit, the probability
distribution is highly intermediate in space and the behavior is completely
dominated by very unlikely events.
\end{abstract}

\maketitle

\section{Introduction}

The last decade has seen increasingly detailed development of the fractional
equations describing the anomalous transport in physics, biology, chemistry
[1--4]. Special attention has been paid to slow subdiffusive transport for
which mean squared displacement is sublinear $<x^{2}(t)>\sim t^{\mu },$
where $\mu $ is the anomalous exponent $\mu <1$. Subdiffusion is
experimentally observed for proteins and lipids on cell membranes \cite{Sa},
RNA molecules in the cells \cite{Gold}, transport in spiny dendrites \cite%
{Santa}, etc. The major feature of this process is the absence of
characteristic microscopic time scale. The theory of anomalous subdiffusion
leads to fractional partial differential equations involving memory effects.
If we introduce the probability density function $p(x,t)$ for finding the
particle in the interval $\left( x,x+dx\right) $ at time $t$, then the
subdiffusive transport of the particles under the influence of external
time-independent force can be described by the fractional Fokker-Planck
(FFP) equation
\begin{equation}
\frac{\partial p}{\partial t}=\mathcal{D}_{t}^{1-\mu }L_{FP}p  \label{FP}
\end{equation}%
with%
\begin{equation}
L_{FP}p=-\frac{\partial \left( v_{\mu }(x)p\right) }{\partial x}+\frac{%
\partial ^{2}\left( D_{\mu }(x)p\right) }{\partial x^{2}}.
\end{equation}
The Riemann-Liouville derivative $\mathcal{D}_{t}^{1-\mu }$ is defined as%
\begin{equation}
\mathcal{D}_{t}^{1-\mu }p\left( x,t\right) =\frac{1}{\Gamma (\mu )}\frac{%
\partial }{\partial t}\int_{0}^{t}\frac{p\left( x,u\right) du}{(t-u)^{1-\mu }%
}
\end{equation}%
and the anomalous exponent $\mu <1$ is assumed to be constant.

The central result of this paper is that the subdiffusive fractional
equations with constant $\mu $ in a bounded domain $\left[ 0,L\right] $ are
not structurally stable with respect to the non-homogeneous variations of
parameter $\mu $. It turns out that the space variations of the anomalous
exponent lead to a drastic change in asymptotic behavior of $p(x,t)$ for
large $t.$ To show this high sensitivity to non-homogeneous perturbations,
one can consider the following exponent
\begin{equation}
\mu (x)=\mu +\delta \nu (x)  \label{per}
\end{equation}%
with constant $\mu $ and perturbation $\delta \nu (x)$ (see Fig. 1). The
asymptotic long-time behavior of the density $p(x,t)$ with (\ref{per}) is
quite different from that of the solution to (\ref{FP}) with the constant
value of $\mu $. It means that the standard subdiffusive equation with
constant $\mu $ is not a robust model for subdiffusive transport in
heterogeneous complex media.
\begin{figure}[tbp]
\begin{center}
\includegraphics[scale=0.35]{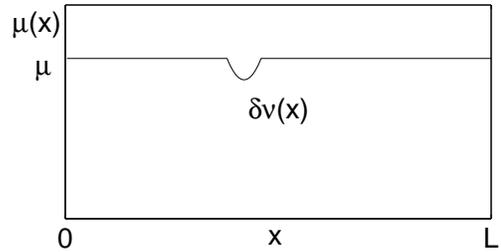}
\end{center}
\par
\label{fig1}
\caption{Non-uniform distribution of anomalous exponent $\protect\mu (x)$ on
the interval $[0,L]$.}
\end{figure}

Now let us explain our main result. The standard way to deal with the
fractional equation like (\ref{FP}) in the bounded domain $\left[ 0,L\right]
$ is a method of separation of variables [1]. Let us consider the case of
the reflecting boundaries at $x=0$ and $x=L$ when (\ref{FP}) has a
stationary solution $p_{st}(x)$ \ satisfying $v_{\mu }(x)p_{st}=\partial
(D_{\mu }(x)p_{st})/\partial x.$ We can write a partial solution of (\ref{FP}%
) as $p\left( x,t\right) =p_{st}(x)Q(x)T(t),$ then the time evolution is
described by fractional relaxation equation
\begin{equation}
\frac{\partial T}{\partial t}=-\lambda \mathcal{D}_{t}^{1-\mu }T,
\end{equation}%
where $\lambda $ is the separation constant. The slow relaxation process
from the initial distribution $p_{0}(x)$ is described by
\begin{equation}
p\left( x,t\right) =p_{st}(x)\sum_{n=0}^{\infty }E_{\mu }\left( -\lambda
_{n}t^{\mu }\right) Q_{n}(x)p_{0n},  \label{as}
\end{equation}%
where $p_{0n}=\int_{0}^{L}p_{0}(x)Q_{n}(x)dx$ and $Q_{n}(x)$ are the eigenfunctions of
\begin{equation}
L_{FP}^{\ast }Q==-\lambda Q.
\end{equation}%
Here the operator $L_{FP}^{\ast }$ is the adjoint of $L_{FP}$
\begin{equation}
L_{FP}^{\ast }Q=:v_{\mu }(x)\frac{\partial Q}{\partial x}+D_{\mu }(x)\frac{%
\partial ^{2}Q}{\partial x^{2}}
\end{equation}
(see details in \cite{Gar,L}). The
only difference between standard Fokker-Planck equation and FFP equation is
the rate of relaxation of $p\left( x,t\right) \rightarrow p_{st}(x)$. In the
anomalous case the relaxation process is very slow and it is described by a
Mittag-Leffler function $E_{\mu }\left( -\lambda _{n}t^{\mu }\right) $ with
the power-law decay $t^{-\mu }$ as $t\rightarrow \infty .$ The exponential
decay $\exp \left( -\lambda _{n}t\right) $ is recovered for $\mu =1$.

In this paper we show that if we consider nonuniform perturbations of
anomalous exponent as (\ref{per}), this relaxation picture is completely
changed. The method of separation of variables does not work for space
dependent $\mu (x).$ The asymptotic behavior of $p\left( x,t\right) $ as $%
t\rightarrow \infty $ is essentially different from that given by (\ref{as}%
). It turns out that in the limit $t\rightarrow \infty $ the probability
density $p\left( x,t\right) $ concentrates around the point $x$, where the
perturbation $\delta \nu (x)$ is located, while the stationary distribution $%
p_{st}(x)$ is completely irrelevant (see Fig. 2 and Fig. 3).

\section{ Fractional master equation with space dependent anomalous exponent}

The question is how to take into account the non-uniform distribution of the
anomalous exponent $\mu $. We cannot simply substitute the expression like (%
\ref{per}) into (\ref{FP}). So we need a fractional master equation with
space dependent $\mu (x).$ Chechkin, Gorenflo and Sokolov were the first to
derive the fractional diffusion equation with varying fractional exponent
\cite{Ch}. They studied a composite system with only two separate regions
with different anomalous exponents and found interesting effects involving
non-trivial average drift. A similar phenomenon has been analyzed in terms
of two equations with different exponent by Korabel and Barkai \cite{KB}.

\subsection{ Hazard function and structured probability density function}

Here we present an alternative derivation which is valid for a general space
and time dependent jump densities. Consider a `space-jump' random walk model
in one space dimension. The particle movement can be described as follows.
It waits for a random time (residence time) $T_{x}$ at each point $x$ in
space before making a jump to another point. The index $x$ indicates that
the waiting time $T_{x}$ depends on a space coordinate $x$. It is convenient
to define the hazard function \cite{Cox} as the escape rate of a walker from
the point $x$
\begin{equation}
\gamma (x,\tau )=\lim_{h\rightarrow 0}\frac{\Pr (\tau <T_{x}<\tau +h\mid
T_{x}>\tau )}{h}.  \label{tran}
\end{equation}%
Next step is the introduction of the structured probability density function
$\xi (x,t,\tau )$ that the particle position $X(t)$ at time $t$ is in the
interval $(x,x+dx)$ and its residence time $T_{x}$ at point $x$ is in the
interval $(\tau ,\tau +d\tau ).$ The advantage of the structured density $%
\xi $ is that a random walk can be considered as Markovian. This is a
standard way to deal with non-Markovian processes \cite{Cox} (see also \cite%
{VR,YH, Fedo}). This density $\xi (x,t,\tau )$ obeys the balance equation
\begin{equation}
\frac{\partial \xi }{\partial t}+\frac{\partial \xi }{\partial \tau }%
=-\gamma (x,\tau )\xi .  \label{basic}
\end{equation}%
Here we consider only the case when the residence time of random walker at $%
t=0$ is equal to zero, so the initial condition is
\begin{equation}
\xi (x,0,\tau )=p_{0}(x)\delta (\tau ),
\end{equation}%
where $p_{0}(x)$ is the density for the initial position $X(0)$. The
boundary condition at $\tau =0$ can be written as \cite{Cox}
\begin{equation}
\xi (x,t,0)=\int_{\mathbb{R}}\int_{0}^{t}\gamma (x,\tau )\xi (x-z,t,\tau
)w(z|x-z,t)d\tau dz,  \label{arr}
\end{equation}%
where $w(z|x,t)$ is the probability density for jumps $z$ from the point $x$
at time $t$ (jumps are independent from the residence time).

Our purpose now is to derive the fractional master equation for the
probability density
\begin{equation}
p(x,t)=\int_{0}^{t}\xi (x,t,\tau )d\tau .  \label{denG}
\end{equation}%
It is convenient to introduce the integral escape rate
\begin{equation}
i(x,t)=\int_{0}^{t}\gamma (\tau ,x)\xi (x,t,\tau )d\tau  \label{bau1}
\end{equation}%
and integral arrival rate
\begin{equation}
j(x,t)=\xi (x,t,0)
\end{equation}%
as the density of particles with zero residence time. The boundary condition
(\ref{arr}) can be rewritten as
\begin{equation}
j\left( x,t\right) =\int_{\mathbb{R}}i\left( x-z,t\right) w\left(
z|x-z,t\right) dz.  \label{ba0}
\end{equation}%
Differentiation of (\ref{denG}) with respect to time and substitution of $%
\partial \xi /\partial t$ from (\ref{basic}) together with (\ref{ba0}) gives
\begin{equation}
\frac{\partial p}{\partial t}=\int_{\mathbb{R}}i\left( x-z,t\right) w\left(
z|x-z,t\right) dz-i(x,t).  \label{Master4}
\end{equation}%
To close this equation we need to express the escape rate $i(x,t)$ in terms
of $p(x,t).$ We solve (\ref{basic}) by the method of characteristics%
\begin{equation}
\xi (x,t,\tau )=\xi (x,t-\tau ,0)e^{-\int_{0}^{\tau }\gamma (x,s)ds},\quad
\tau <t.  \label{s1}
\end{equation}%
Here we recognize the survival function \cite{Cox}%
\begin{equation}
\Psi (x,\tau )=\Pr \left\{ T_{x}>\tau \right\} =e^{-\int_{0}^{\tau }\gamma
(x,s)ds}
\end{equation}%
so the structural density $\xi $ can be rewritten as%
\begin{equation}
\xi (x,t,\tau )=j\left( x,t-\tau \right) \Psi (x,\tau ),\quad \tau <t.
\end{equation}%
The residence time PDF $\phi (x,\tau )$ is related to $\gamma (x,\tau )$\ as
\begin{equation}
\phi (x,\tau )=-\partial \Psi /\partial \tau =\gamma (x,\tau )\exp \left(
-\int_{0}^{\tau }\gamma (x,s)ds\right) .
\end{equation}%
The balance equation for $p\left( x,t\right) $ can be found by substitution
of (\ref{s1}) and the initial condition $\xi (x,0,\tau )=p_{0}(x)\delta
(\tau )$ into (\ref{denG})
\begin{equation}
p\left( x,t\right) =\int_{0}^{t}j\left( x,u\right) \Psi
(x,t-u)du+p_{0}\left( x\right) \Psi (x,t).  \label{ba1}
\end{equation}%
To obtain the equation for $i(x,t)$ we substitute (\ref{s1}) and the initial
condition into (\ref{bau1})
\begin{equation}
i(x,t)=\int_{0}^{t}j(x,u)\phi (x,t-u)du+p_{0}(x)\phi (x,t).  \label{ba00}
\end{equation}%
Using the Laplace transform in (\ref{ba1}) and (\ref{ba00}) we eliminate $%
j(x,t)$ and obtain \cite{Ch}
\begin{equation}
i\left( x,t\right) =\int_{0}^{t}K\left( x,t-\tau \right) p\left( x,\tau
\right) d\tau ,  \label{i}
\end{equation}%
where $K(x,t)$ is the memory kernel defined by its Laplace transform
\begin{equation}
\hat{K}\left( x,s\right) =\frac{\hat{\phi}\left( x,s\right) }{\hat{\Psi}%
\left( x,s\right) }.
\end{equation}

\subsection{ Anomalous subdiffusion in heterogeneous media}

Let us consider the anomalous subdiffusive case with the survival
probability \cite{Scalas}
\begin{equation}
\Psi \left( x,t\right) =E_{\mu (x)}\left[ -\left( \frac{t}{\tau (x)}\right)
^{\mu (x)}\right] ,\ 0<\mu (x)<1,  \label{ML}
\end{equation}%
where $E_{\mu }\left[ z\right] $ is the Mittag-Leffler function. The Laplace
transforms of $\Psi \left( x,t\right) $ and $\phi (x,t)$ are
\begin{equation}
\hat{\Psi}\left( x,s\right) =\frac{\tau (x)\left( s\tau (x)\right) ^{\mu
(x)-1}}{1+\left( s\tau (x)\right) ^{\mu (x)}},\ \hat{\phi}\left( x,s\right) =%
\frac{1}{1+\left( s\tau (x)\right) ^{\mu (x)}}.
\end{equation}%
The Laplace transform of the memory kernel $K\left( x,t\right) $ is%
\begin{equation}
\hat{K}\left( x,s\right) =\frac{s^{1-\mu (x)}}{\tau (x)^{\mu (x)}}
\label{KK}
\end{equation}%
and the integral escape rate $i\left( x,t\right)$ can be written as 
\begin{equation}
i\left( x,t\right) =\frac{1}{\tau (x)^{\mu (x)}}\mathcal{D}_{t}^{1-\mu (x)}p\left( x,t\right).
\end{equation}%
Substitution of this expression into (\ref{Master4}) gives the
fractional master equation
\begin{eqnarray}
\frac{\partial p}{\partial t} &=&\int_{\mathbb{R}}\frac{\mathcal{D}%
_{t}^{1-\mu (x-z)}p\left( x-z,t\right) }{\tau (x-z)^{\mu (x-z)}}w\left(
z|x-z,t\right) dz  \notag \\
&&-\frac{1}{\tau (x)^{\mu (x)}}\mathcal{D}_{t}^{1-\mu (x)}p\left( x,t\right)
,  \label{Master8}
\end{eqnarray}%
where $\mathcal{D}_{t}^{1-\mu (x)}$ is the Riemann-Liouville fractional
derivative with varying order. This equation can be used to derive the
general Fokker-Planck equation \cite{HLS}. If we assume that the anomalous
exponent $\mu $ and time parameter $\tau $ are independent from coordinate $x
$, this equation can be rewritten in terms of Caputo derivative%
\begin{equation}
\tau ^{\mu }\frac{\partial ^{\mu }p}{\partial t^{\mu }}=\int_{\mathbb{R}%
}p\left( x-z,t\right) w\left( z|x-z,t\right) dz-p(x,t).
\end{equation}%
It should be noted that the fractional equation with the Caputo derivative
cannot be served as a model for subdiffusion in heterogeneous media with
varying in space anomalous exponent $\mu (x)$. \

Master equation (\ref{Master8}) can be a starting point for deriving
nonlinear fractional equations. If instead of $p$ we consider the mean
density of particles $\rho $ and assume that jump PDF $w\left( z\right) $
depends on $\rho ,$ then one can write
\begin{eqnarray}
\frac{\partial \rho }{\partial t} &=&\int_{\mathbb{R}}\frac{\mathcal{D}%
_{t}^{1-\mu (x-z)}\rho \left( x-z,t\right) }{\tau (x-z)^{\mu (x-z)}}w\left(
z|\rho (x-z,t\right) )dz  \notag \\
&&-\frac{1}{\tau (x)^{\mu (x)}}\mathcal{D}_{t}^{1-\mu (x)}\rho \left(
x,t\right) .  \label{Master99}
\end{eqnarray}%
Expansion of this equation in $z$ can give a variety of fractional
non-linear PDE's. As an example, let us consider the case of the symmetrical
kernel $w(z|\rho )$ for which the first moment $\int_{\mathbb{R}}zw\left(
z|\rho (x,t\right) dz=0.$ Then (\ref{Master99}) can be approximated by a
non-linear fractional equation
\begin{equation}
\frac{\partial \rho }{\partial t}=\frac{\partial ^{2}\left( D_{\mu }(\rho )%
\mathcal{D}_{t}^{1-\mu (x)}\rho \right) }{\partial x^{2}}
\end{equation}%
with varying anomalous exponent $\mu (x)$ and nonlinear fractional diffusion
coefficient $D_{\mu }(\rho )$ $:$
\begin{equation}
D_{\mu }(\rho )=\frac{m_{2}\left( \rho \right) }{2\tau (x)_{{}}^{\mu (x)}}%
,\quad m_{2}\left( \rho \right) =\int_{\mathbb{R}}z^{2}w\left( z|\rho
\right) dz.
\end{equation}

First, let us consider random walk on a lattice with the space size $a.$ We
denote the probability of a particle moving right and left from the point $x$
as $r(x)$ and $l(x)$ correspondingly ($r(x)+l(x)=1$). Then the jump pdf can
be written as%
\begin{equation}
w\left( z|x\right) =r(x)\delta (z-a)+l(x)\delta (z+a).
\end{equation}%
The fractional master equation (\ref{Master8}) takes the form
\begin{eqnarray}
\frac{\partial p}{\partial t} &=&\frac{r(x-a)}{\tau (x-a)^{\mu (x-a)}}%
\mathcal{D}_{t}^{1-\mu (x-a)}p\left( x-a,t\right) +  \notag \\
&&\frac{l(x+a)}{\tau (x+a)^{\mu (x+a)}}\mathcal{D}_{t}^{1-\mu (x+a)}p\left(
x+a,t\right) -  \notag \\
&&\frac{1}{\tau (x)^{\mu (x)}}\mathcal{D}_{t}^{1-\mu (x)}p\left( x,t\right) .
\label{walk}
\end{eqnarray}%
In the limit of small $a$ and $\tau (x)$ \cite{BMK} one can obtain from (\ref%
{walk}) the FFP equation with varying anomalous exponent
\begin{equation}
\frac{\partial p}{\partial t}=-\frac{\partial \left( v_{\mu }(x)\mathcal{D}%
_{t}^{1-\mu (x)}p\right) }{\partial x}+\frac{\partial ^{2}\left( D_{\mu }(x)%
\mathcal{D}_{t}^{1-\mu (x)}p\right) }{\partial x^{2}}  \label{FPnew}
\end{equation}%
with the finite values of the fractional diffusion coefficient $D_{\mu }(x)$
and fractional drift $v_{\mu }(x):$
\begin{equation}
D_{\mu }(x)=\frac{a^{2}}{2\tau (x)_{{}}^{\mu (x)}},\quad v_{\mu }(x)=\frac{%
2(r(x)-l(x))D_{\mu }(x)}{a}.
\end{equation}%
Note that in order to keep the fractional drift $v_{\mu }(x)$ finite as $%
a\rightarrow 0$, we need to assume that $r(x)-l(x)=O(a).$

If we put the reflecting barriers at $x=0$ and $x=L$ and consider constant
exponent $\mu $ and diffusion $D_{\mu },$ then the FFP equation (\ref{FPnew}%
) admits the stationary solution in the form of the Gibbs-Boltzmann
distribution%
\begin{equation}
p_{st}(x)=C\exp \left[ -U(x)\right] ,\quad U(x)=-\frac{1}{D_{\mu }^{{}}}%
\int^{x}v_{\mu }(z)dz  \label{sta}
\end{equation}%
with $C^{-1}=\int_{0}^{L}\exp \left[ -U(x)\right] dx$.

If $\mu $ is constant, the fractional time derivative does not affect the
Gibbs-Boltzmann distribution \cite{MK1,DCS}. But this result is structurally
unstable with respect to any non-uniform variations of $\mu .$ Let us show
now that the Gibbs-Boltzmann distribution (\ref{sta}) is absolutely
irrelevant for the long time behavior of the solution to the FFP equation (%
\ref{FPnew}) with non-uniform distribution of $\mu (x)$ (\ref{per}).

\subsection{Discrete model}

We divide the interval $[0,L]$ into $n$ discrete states. At each state $i$,
the probability of jumping in the neighborhood to the left or right is given
respectively by $l_{i}$ and $r_{i}$ ($l_{i}+r_{i}=1$). The fractional
equation (\ref{walk}) for $p_{i}(t)=\Pr \left\{ X(t)=i\right\} $ can be
rewritten as
\begin{eqnarray}
p_{i}^{\prime }(t) &=&\frac{r_{i-1}\mathcal{D}_{t}^{1-\mu _{i-1}}p_{i-1}(t)}{%
\tau _{i-1}{}^{\mu _{i-1}}}+\frac{l_{i+1}\mathcal{D}_{t}^{1-\mu
_{i+1}}p_{i+1}(t)}{\tau _{i+1}{}^{\mu _{i+1}}}  \notag \\
&&-\frac{\mathcal{D}_{t}^{1-\mu _{i-1}}p_{i}(t)}{\tau _{i}{}^{\mu _{i}}}%
,\quad i=1,\ldots ,n  \label{dis}
\end{eqnarray}%
subject to the conditions $l_{1}=r_{-1}=0,$ $r_{1}=1$ and $l_{n}=1,$ $%
r_{n}=l_{n+1}=0$. Note that the FFP equation (\ref{FPnew}) is just a
continuous approximation of Eq. (\ref{dis}). Taking the Laplace transform of
(\ref{dis}) and using $\sum_{i}\hat{p}_{i}(s)=\frac{1}{s}$, we obtain
\begin{eqnarray}
&&s\hat{p}_{i}(s)\left( 1+\frac{r_{i-1}}{(s\tau _{i-1}{})^{\mu _{i-1}}}+%
\frac{l_{i+1}}{(s\tau _{i+1}{})^{\mu _{i+1}}}+\frac{1}{(s\tau _{i}{})^{\mu
_{i}}}\right)   \notag \\
&=&\frac{r_{i-1}}{(s\tau _{i-1}^{{}})^{\mu _{i-1}}}\left( 1-\sum_{j\neq
i-1,i}s\hat{p}_{j}(s)\right)   \notag \\
&&+\frac{l_{i+1}}{(s\tau _{i+1}{})^{\mu _{i+1}}}\left( 1-\sum_{j\neq i,i+1}s%
\hat{p}_{j}(s)\right) +p_{i}(0)
\end{eqnarray}%
If one $\mu _{M}$ is smaller than the others ($\mu _{M}<\mu _{i}$ $\forall i)
$, one can find that $s\hat{p}_{i}(s)\rightarrow 0$ and $s\hat{p}%
_{M}(s)\rightarrow 1$ as $s\rightarrow 0.$ It means that in the limit $%
t\rightarrow \infty ,$ we obtain
\begin{equation}
p_{i}(t)\rightarrow 0,\qquad p_{M}(t)\rightarrow 1.
\end{equation}%
This result in a continuous case can be rewritten as $p\left( x,t\right)
\rightarrow \delta (x-x_{\min })$ as $t\rightarrow \infty ,$ where $x_{\min }
$ is the point on the interval $\left[ 0,L\right] $ at which $\mu (x)$ takes
its minimum value. A similar result was obtained for a symmetrical random
walk in \cite{Fedo} in the context of chemotaxis (anomalous aggregation).
Note that Shushin \cite{Shushin} considered a two-state anomalous system
with different anomalous exponent $\mu $ and found that in the long time
limit the probability is located in the slower state (see also \cite{CFM,KB}%
).

\section{Monte Carlo simulations}

To validate our results, we run Monte Carlo simulations with the following
procedure. Random numbers with uniform distribution, $u$ and $v$, are
generated and then transformed into Mittag-Leffler distributed random
numbers using the following inversion formula $t_{\mu }=-\tau \log (u)\left(
\frac{\sin (\mu \pi )}{\tan (\mu \pi v)}-\cos (\mu \pi )\right) ^{\frac{1}{%
\mu }}$ \cite{Ra} (see for details \cite{Monte}). We \ take $L=1$ and divide
the interval $[0,1]$ into $100$ subintervals. We use $r_{i}=1/2+5a(1-2ai)/2,$
$1\leq i\leq 100$ and $a=1/100.$ This corresponds to
\begin{equation}
r(x)=1/2+5a(1/2-x),
\end{equation}%
so the drift $v_{\mu }(x)=10(1-2x)D_{\mu }$ and the potential $%
U(x)=5(1-2x)^{2}/2.$ All the random walkers start in the same state $i=40$,
their number $N=10^{4}$, $\tau _{i}=10^{-4}$ for all $i$, and the long time
limit is set at $T=10^{5}.$

First step is to compute the exact stationary PDF given by (\ref{sta}) and
see how well our Monte Carlo simulations work. Fig. 2 shows that the Monte
Carlo simulations agree with the Gibbs-Boltzmann distribution.
\begin{figure}[tbp]
\begin{center}
\includegraphics[scale=0.35]{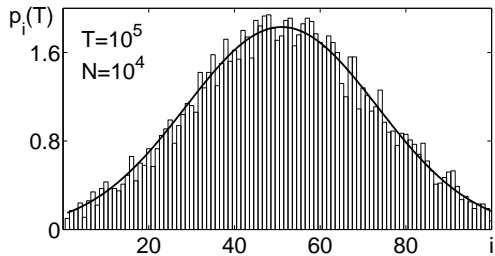} \label{fig2}
\end{center}
\caption{Long time limit of the solution to the system (\protect\ref{dis})
with $\protect\mu _{i}=0.5$ for all $i$. Gibbs-Boltzmann distribution is
represented by the line. }
\end{figure}
\begin{figure}[tbp]
\begin{center}
\includegraphics[scale=0.35]{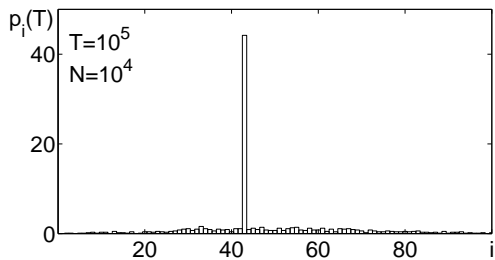} \label{fig3}
\end{center}
\caption{Long time limit of the system (\protect\ref{dis}) when $\protect\mu %
_{i}$ is subject to a perturbation. The parameters are $\protect\mu _{i}=0.5
$ for all $i$ except $i=42$ for which $\protect\mu _{42}=0.3$.}
\end{figure}

The next step is to show that the Gibbs-Boltzmann distribution (\ref{sta})
is absolutely irrelevant as far as the long time behavior of non-uniform
system is concerned. The anomalous exponent $\mu _{i}$ is assumed to be $0.5$
for all states except one, $i=42,$ for which $\mu _{42}=0.3$. One can see
from Fig. 3 that in the long time limit the probability is concentrated at
state $i=42$. One can conclude that there is a complete break down in the
predictions based on the FFP equation with uniform anomalous exponent. If
the system was structurally stable we would expect to see something more
like Fig. 2 again. However, the outcome is completely dominated by the
perturbation $\mu _{42}=0.3$. This result has a huge implication for
modelling anomalous subdiffusive transport of proteins, porous media, etc.
In reality the environment in which anomalous transport takes place is never
homogeneous.

Several attempts have been made to take into account the random distribution
of anomalous exponent (see, for example, \cite{CKS,M}). One can introduce
PDF $f\left( \mu \right) $ for a random $\mu $ and write down the
distributed-order fractional FPE as
\begin{equation}
\int_{0}^{1}\tau ^{\mu -1}\frac{\partial ^{\mu }p}{\partial t^{\mu }}f\left(
\mu \right) d\mu =L_{FP}p.  \label{order}
\end{equation}%
Let us show that if we generate the random field $\mu (x)$ along the space
interval $[0,1],$ the asymptotic behavior of $p(x,t)$ will be quite
different from that of the average fractional equation (\ref{order}).
\begin{figure}[tbp]
\begin{center}
\includegraphics[scale=0.35]{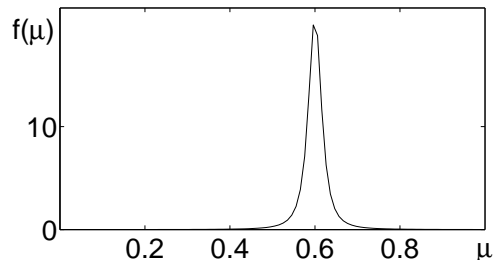} \label{fig4}
\end{center}
\caption{The PDF $f\left( \protect\mu \right) $ of random anomalous exponent
$\protect\mu $. }
\end{figure}

\begin{figure}[tbp]
\begin{center}
\includegraphics[scale=0.35]{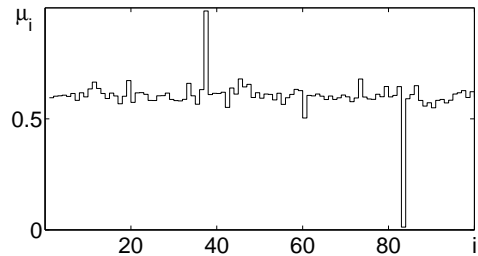} \label{fig5}
\end{center}
\caption{One sample of the discrete random field $\protect\mu_i$ along $i$
for $1\le i\le 100$. }
\end{figure}

\begin{figure}[tbp]
\begin{center}
\includegraphics[scale=0.35]{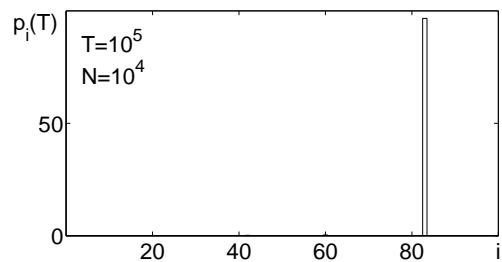} \label{fig6}
\end{center}
\caption{Long time limit of the system (\protect\ref{dis}) when $\protect\mu %
_{i}$ is the random field represented in Fig. 5.}
\end{figure}
Fig. 4 shows the PDF $f\left( \mu \right) $ which will be used to generate
the discrete uncorrelated random field $\mu _{i}$. The probability is
concentrated around the point $0.6$ such that $\Pr \left\{ 0.5<\mu
<0.7\right\} =0.98$. This distribution is chosen so that extreme values are
highly unlikely to occur, with a purpose to show that the extreme low values
dominate the long time behavior. Fig. 5 shows one sample of random field $%
\mu (x)$ on the interval $[0,1]$ which is subdivided into $100$ subintervals
($1\leq i\leq 100$). Fig. 5 shows clearly that the values of $\mu _{i}$
fluctuate around the mean. The value at $\mu _{82}=0.01245$ has a very small
probability, since$~\Pr \left\{ \mu <0.02\right\} =2.5\times 10^{-4}$. It is
a very unlikely event, yet one can see from Fig. 6 the state $i=82$
completely dominates the long time outcome of (\ref{dis}). This phenomenon
can be interpreted as a `Black Swan'. The distribution of $p(x,t)$ is highly
intermediate for large $t$, so the average behavior described by (\ref{order}%
) can be very misleading. It has been found \cite{CKS} that the distribution
of the anomalous exponent in Eq.(\ref{order}) leads to ultra-slow kinetics,
but the stationary distribution is still given by the Gibbs-Boltzmann
distribution \cite{DCS}. Our results show that random space variation of the
anomalous exponent leads to completely different behavior in the long time
limit (see Fig. 6). It should be noted that anomalous diffusion is just an intermediate asymptotic. 
When time tends to infinity we expect a cross-over from anomalous diffusion to normal diffusion, 
and then we will recover the Gibbs-Boltzmann distribution.

The standard tool for studying a subdiffusion is a subordination technique
\cite{Weron} with constant anomalous exponent. It would be interesting to
apply similar technique if possible to non-homogeneous case. It would be
also interesting to take into account chemical reactions together with
non-uniform anomalous exponent \cite{Fedo2}.

\section{ Conclusions}

We have demonstrated that when the anomalous exponent $\mu $ depends on the
space variable $x$, the Gibbs-Boltzmann distribution is not a long time
limit of the fractional Fokker-Planck equation. Even very small variations
of the exponent lead to a drastic change of $p(x,t)$ in the limit $%
t\rightarrow \infty $. We have derived the fractional master equation with
space dependent anomalous exponent. We analyzed asymptotic behavior of
corresponding lattice model in a finite domain with $n$ states with
different exponents. We have found that in this situation the probabilities\
$p_{i}(t)$ do not converge to the stationary distribution.\ To illustrate
our ideas, we ran Monte Carlo simulations which show a complete break down
in the predictions based on the FFP equation with uniform anomalous
exponent. Further, we have shown that the idea of taking into account the
randomness of anomalous exponent $\mu $ by averaging the fractional equation
with respect to the distribution $f(\mu )$ is not applicable to a
non-homogeneous finite domain. Monte Carlo simulations show that for every
random realization of $\mu (x)$ the PDF $p(x,t)$ is highly intermediate, so
the average behavior can be misleading. Although it is possible in theory to
have a completely homogeneous environment, in which $\mu $ is uniform, it is
not useful in any real application like chemotaxis \cite{Fedo} or morphogen
gradient formation \cite{Yuste} because any non-homogeneous variation
destroys the predictions based on this model in the long time limit.

\end{document}